\documentstyle[emulateapj]{article}

\lefthead{Nagao et al.}
\righthead{Advection Dominated Accretion Flow in Radio Galaxies}

\begin{document}

\submitted{To Appear in The Astrophysical Journal} 
\title{Is There an Advection Dominated Accretion Flow
       in Radio Galaxies with Double-Peaked Balmer Lines ?}

\author{Tohru NAGAO, Takashi MURAYAMA, Yasuhiro SHIOYA, 
        and Yoshiaki TANIGUCHI}
\affil{Astronomical Institute, Graduate School of Science, 
       Tohoku University, Aramaki, Aoba, Sendai 980-8578, Japan}

\begin{abstract}

In order to examine the prediction that the broad-line radio galaxies 
(BLRGs) with double-peaked Balmer lines harbor an accretion disk 
characterized by an 
advection-dominated accretion flow (ADAF) in their nuclei, we investigate 
narrow emission-line flux ratios 
of the narrow-line regions which are photoionized by the nuclear 
continuum radiation. We compile data from the literature and
confirm the pioneering work of Eracleous \& Halpern that the
BLRGs with the double-peaked Balmer emission exhibit larger flux ratios of
both [O {\sc i}]$\lambda$6300/[O {\sc iii}]$\lambda$5007 and 
[O {\sc ii}]$\lambda$3727/[O {\sc iii}]$\lambda$5007 than the
BLRGs without the double-peaked Balmer emission.
To examine whether or not these properties are attributed to 
the difference in the spectral energy distribution (SED) of the 
ionizing radiation between the BLRGs with and without
the double-peaked Balmer emission,
we perform photoionization model calculations using two types of input 
continuum radiation; one has the strong big blue bump which is expected 
for standard optically-thick accretion disks and
another does not exhibit a strong big blue bump as expected for 
optically-thin ADAFs.
We find that the data of the BLRGs with the double-peaked Balmer lines
are consistent with the models adopting the SED without a strong big blue 
bump while the data of the BLRGs without the double-peaked emission lines
are well described by the models adopting the SED with a strong big blue bump.
On the other hand, the observed differences in the NLR emission is hard to be
explain by the difference in the contribution of shocks.
These results support the idea that the double-peaked Balmer lines arise
at an outer region of an accretion disk which is illuminated by an inner,
geometrically-thick ADAF.

\end{abstract}


\section{INTRODUCTION}

\begin{figure*}
\epsscale{0.535}
\plotone{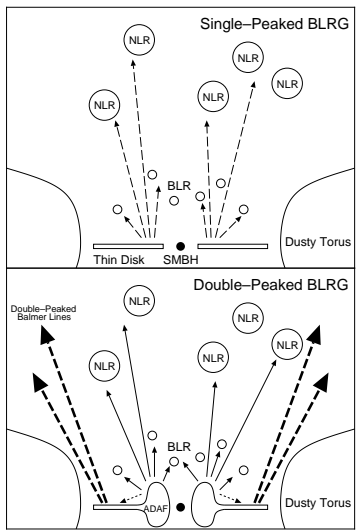}
\caption{
(Top) Cartoon of the nuclear region in BLRGs without double-peaked
   Balmer lines. A supermassive black hole is surrounded by a
   standard geometrically-thin, optically-thick accretion disk,
   whose inner part radiates thermal UV/soft X-ray emission (dashed 
   black arrows). This thermal emission ionizes gas clouds in the
   broad-line region (BLR) and the narrow-line region (NLR), which are shown 
   by small and large circles, respectively.
   Because of a low temperature, an outer part of this standard disk
   does not radiate ionizing photons and permitted lines effectively.
(Bottom) Cartoon of the nuclear region in BLRGs with double-peaked
   Balmer lines. The inner region of the standard thin disk is replaced by
   a puffed-up structure, i.e., an ADAF. The ADAF radiates a relatively 
   hard continuum, which ionizes gas in the surrounding 
   BLR and NLR (solid black arrows). A part of this hard continuum emission
   irradiates the outer thin disk (dotted arrows), from which the
   double-peaked Balmer lines arise (dashed thick arrows).
   Note that the observed profile of Balmer lines of 
   double-peaked Balmer lines should consist of
   the three components in this scheme; a double-peaked broad component,
   a single-peaked broad component, and a narrow component.
   This is consistent with some observations (e.g., Halpern et al. 1996).
\label{fig1}}
\end{figure*}

It has been reported that double-peaked broad Balmer lines are seen
in the optical spectra of some broad-line radio galaxies (BLRGs)
[e.g., Osterbrock, Koski, \& Phillips 1976; Halpern \& Filippenko 1988; 
P\'{e}rez et al. 1988; Halpern 1990; Veilleux \& Zheng 1991; 
Colina, L\'{\i}pari, \& Macchetto 1991;
Eracleous \& Halpern 1994 (EH94); Halpern \& Eracleous 1994; 
Sulentic et al. 1995; Halpern et al. 1996] and
LINERs (e.g., Storchi-Bergmann, Baldwin, \& Wilson 1993; 
Bower et al. 1996; Ho et al. 2000; Shields et al. 2001).
Since this remarkable feature may contain kinematical and geometrical
information concerning the ionized gas in broad-line regions (BLRs),
various models for the double-peaked emission line have been proposed
up to now. Among such models, the disk-BLR model, in which the
double-peaked profile is explained taking account
of accretion disks as an emitter of broad emission lines, 
has been often investigated since P\'{e}rez et al. (1988) reported
that the profile of the H$\beta$ emission of 3C 390.3 can be well fitted 
by a relativistic accretion disk model [see also 
Chen, Halpern, \& Filippenko 1989 (CHF89); Chen \& Halpern 1989 (CH89)].
However, the gravitational energy released in a standard, geometrically-thin
and optically-thick accretion disk is insufficient to account for
the luminosity of the BLR emission (CHF89; EH94).
Moreover, if the standard disk is assumed, the effective temperature 
in the region where the BLR emission 
is expected to arise is calculated to be less than 5000 K, which is
too low in BLRs (CHF89; see also Collin-Souffrin 1987).
Therefore, an external heating source is required for an accretion disk to be
a source of the BLR emission.

When the mass accretion rate is small compared to the Eddington value,
ions in the inner disk become very hot because of lack of an efficient
cooling process.
The resulting high pressure produces puffed-up structure
with nearly spherical inflow, which is generally called 
advection-dominated accretion flow (ADAF; e.g., Ichimaru 1977; 
Rees et al. 1982; Narayan \& Yi 1995).
CHF89 pointed out that this geometrically-thick ADAF can illuminate
an outer disk that remains geometrically thin (see also CH89;
Shields et al. 2001).
This supplied energy enables the outer thin disk to radiate the BLR 
emission, giving rise to the double-peaked broad Balmer lines.
When the mass accretion rate is high, on the contrary,
we see only single-peaked (``normal'') Balmer lines which may arise in
more distant regions from the nucleus compared to an accretion disk\footnote{
   Eracleous \& Halpern (1994) found that the BLRGs with double-peaked 
   emission lines tend to have larger line width (both FWHM and FWZI) than 
   those without double-peaked emission lines. This suggests that the 
   double-peaked emission lines originate closer to the central black hole 
   than the single-peaked broad emission lines. However, the configuration of
   the gas clouds emitting the single-peaked broad emission lines is not well
   understood. For instance, Rokaki, Boisson, \& Collin-Souffrin (1992)
   reported that even the single-peaked broad emission lines can be fitted 
   by some appropriate disk models (see also Corbin 1997). On the other hand,
   Chiang \& Murray (1996) pointed out that some parts of single-peaked broad
   emission may come from the winds emanating from accretion disks (see also
   Murray \& Chiang 1997).
}.These two situations are shown in Figure 1.
This scheme seems interesting because it explains why BLRGs do not
always show the double-peaked Balmer lines.

Since the ADAFs are optically thin, they do not exhibit a strong 
big blue bump (BBB), i.e., a thermal blue/UV component, in their spectra.
Therefore, the above scheme (hereafter ``illuminated-disk model'') predicts
that an evident difference in the spectral energy distribution (SED) 
in the wavelength of UV to X-ray is expected between the BLRGs with and 
without the double-peaked Balmer lines (e.g., EH94).
Indeed, CHF89 reported that there is no evidence for the strong BBB in
Arp 102B, which is a typical example of BLRGs with 
double-peaked Balmer lines (see also Edelson \& Malkan 1986).
However, since most of the BLRGs are rather faint, the difference in
the SED is hard to be observed directly, except for a few exceptions.
EH94 mentioned that the difference in the SED between standard thin disks
and ADAFs may affect the physical properties of ionized gas in narrow-line
regions (NLRs) which are photoionized by the nuclear continuum radiation. 
They found that the BLRGs with double-peaked Balmer lines 
exhibit stronger [O {\sc i}]$\lambda$6300 emission, which arises at NLRs, 
than those without double-peaked Balmer lines. 
This can be interpreted as the effect of
the harder SED of ADAFs (see also Halpern \& Eracleous 1994;
Halpern 1999; Ho et al. 2000).
However, there is no study to examine this interpretation quantitatively
based on photoionization models. 
Therefore, we investigate the properties of the NLRs in the BLRGs 
with/without the double-peaked Balmer lines in this paper
to examine whether or not the observed difference in the narrow 
emission-line flux ratios is owing to the difference in the SED.
This attempt seems crucially important toward the understanding of 
not only the origin of the double-peaked emission lines but also 
the configuration of BLRs and accretion disks in active galactic nuclei
(AGNs).

\section{DATA}

\begin{figure*}
\epsscale{1.2}
\plotone{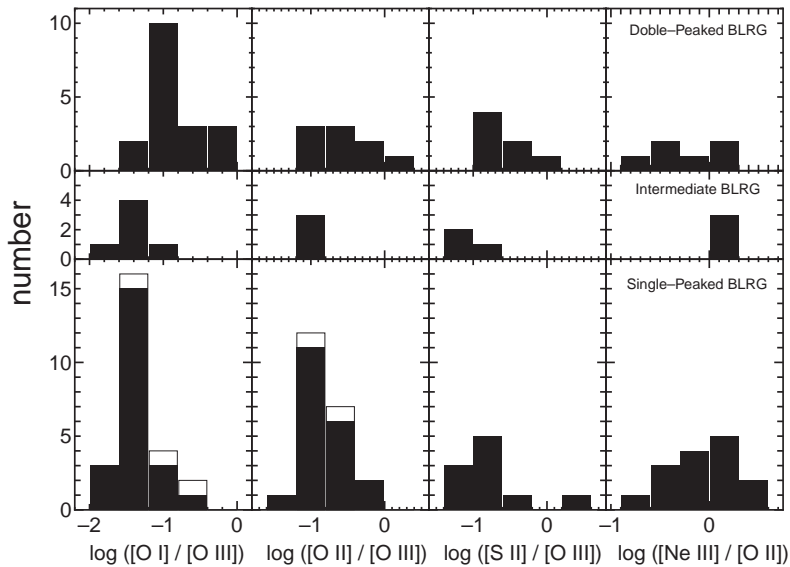}
\caption{
Frequency distributions of the emission-line flux ratios of
[O {\sc i}]$\lambda$6300/[O {\sc iii}]$\lambda$5007,
[O {\sc ii}]$\lambda$3727/[O {\sc iii}]$\lambda$5007,
[S {\sc ii}]$\lambda \lambda$6717,6731/[O {\sc iii}]$\lambda$5007, and
[Ne {\sc iii}]$\lambda$3869/[O {\sc ii}]$\lambda$3727, for
the double-peaked BLRGs (top), the intermediate BLRGs (middle), and 
the single-peaked BLRGs (bottom).
The open boxes denote the upper-limit data.
\label{fig2}}
\end{figure*}

 \subsection{Classification}

EH94 classified 94 BLRGs into five classes based on the profile of
the H$\alpha$ emission as follows;
(1) the BLRGs with double-peaked H$\alpha$ emission whose blue peak
    is stronger than the red peak and whose profile can be fitted
    by their relativistic disk models (12 BLRGs),
(2) the BLRGs with double-peaked H$\alpha$ emission whose blue peak
    is stronger than that of red peak but whose profile cannot be fitted 
    by their relativistic disk models (4 BLRGs),
(3) the BLRGs with double-peaked H$\alpha$ emission whose red peak is
    stronger than the blue peak (6 BLRGs),
(4) the BLRGs without double-peaked H$\alpha$ emission but their
    profile show a pronounced asymmetry or a single displaced peak (9 BLRGs),
and
(5) the BLRGs with normal single-peaked H$\alpha$ emission (63 BLRGs).

Here it should be noted that EH94 adopted simple relativistic disk models 
described by CHF89 and CH89 to fit the observed H$\alpha$ profiles.
Since these models cannot explain the observed double-peaked profiles 
whose red peak is stronger than the blue peak, 
EH94 treated only BLRGs categorized in the
class of (1) as ``disk-like emitters''. EH94 compared the statistical
properties of the 12 disk-like emitters with those of the other 82 BLRGs 
in order to investigate the origin of the double-peaked Balmer lines.
However, it is now recognized that some modified relativistic disk models 
can successfully reconstruct the profiles of double-peaked Balmer lines 
which cannot be explained by the models of CHF89 and CH89; e.g., 
disk plus hot patch models (Veilleux \& Zheng 1991; 
Zheng, Veilleux, \& Grandi 1991), accretion disk with two-arm spiral waves 
(Chakrabarti \& Wiita 1994), and elliptical disk models
(Eracleous et al. 1995; see also Syer \& Clarke 1992).
Therefore, we treat the first three classes of EH94 as
``double-peaked BLRG'' regardless of the peak strength ratio.
These objects are expected to harbor a geometrically-thick ADAF
in their nuclei if the illuminated-disk model is the case.
Then, we call the class (4) of EH94 as ``intermediate BLRG''.
It is not clear whether or not the objects in this class have
the ADAFs and the illuminated thin disk in their nucleus.
And finally, we call the BLRG with normal single-peaked H$\alpha$ emission
as ``single-peaked BLRG''. 
The objects in this class are expected not to
have an evolved geometrically-thick ADAF.
Note that a BLRG, Pictor A, is classified as a double-peak BLRG
in this paper though it was classified as (5) by EH94, 
following some recent reports
that Pictor A also has double-peaked Balmer emission
(Halpern \& Eracleous 1994; Sulentic et al. 1995).
In summary, the EH94 sample is divided into three groups:
23 double-peaked BLRGs, 9 intermediate-type BLRGs, and 62 single-peaked BLRGs.

 \subsection{Data Compilation}

In order to investigate the properties of ionized gas in NLRs for each 
group of BLRGs, we compiled data of forbidden emission-line flux ratios
of the EH94 sample from the literature.
The compiled emission-line flux ratios are
[O {\sc i}]$\lambda$6300/[O {\sc iii}]$\lambda$5007,
[O {\sc ii}]$\lambda$3727/[O {\sc iii}]$\lambda$5007,
[S {\sc ii}]$\lambda \lambda$6717,6731/[O {\sc iii}]$\lambda$5007, and
[Ne {\sc iii}]$\lambda$3869/[O {\sc ii}]$\lambda$3727,
which are given in Table 1.
For the case that a certain emission-line flux ratio of a certain
object is given by more than one paper, the averaged value is given
in this table.
Here we do not use Balmer lines, because the narrow Balmer components
of broad-line objects are often hard to measure correctly (see, e.g.,
Nagao, Murayama, \& Taniguchi 2001c). For the same reason,
estimating the amount of extinction by the Balmer decrement method 
(e.g., Osterbrock 1989) cannot be adopted. 
We cannot use other methods (e.g., using the flux ratio of 
[S {\sc ii}]$\lambda$4071/[S {\sc ii}]$\lambda$10320)
because of small number of available emission-line data.
Therefore, we do not make the reddening correction for the data.
Note that the dust reddening of NLR emission of broad-line objects
are small in general (e.g., De Zotti \& Gaskell 1985;
Dahari \& De Robertis 1988).
The data presented in Table 1 are not corrected for Galactic reddening.
The effect of dust extinction on our results is discussed in section 4.2.

\section{THE NARROW EMISSION-LINE FLUX RATIOS}

\begin{figure*}
\epsscale{1.0}
\plotone{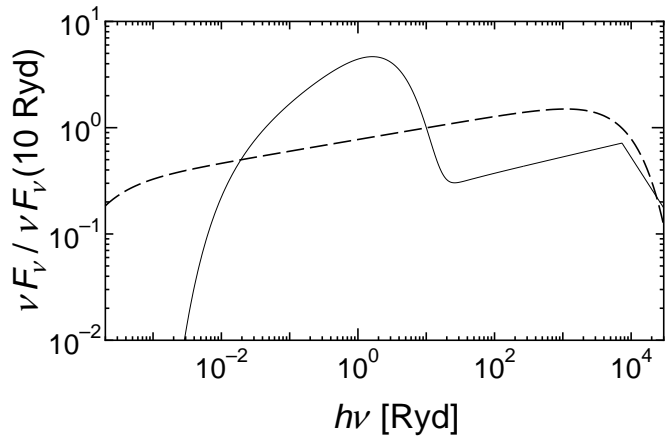}
\caption{
Template SEDs adopted for the photoionization model calculations.
The solid line denotes the ``SED with BBB'' which represents the continuum
emission arising from geometrically-thin and optically-thick accretion
disks, and the dashed line denotes the ``SED without BBB''
which represents the continuum emission arising from 
geometrically-thick and optically-thin ADAFs.
These are normalized at 10 Ryd.
\label{fig3}}
\end{figure*}

In Figure 2, we show the frequency distributions of the emission-line 
flux ratios of
[O {\sc i}]$\lambda$6300/[O {\sc iii}]$\lambda$5007,
[O {\sc ii}]$\lambda$3727/[O {\sc iii}]$\lambda$5007,
[S {\sc ii}]$\lambda \lambda$6717,6731/[O {\sc iii}]$\lambda$5007, and
[Ne {\sc iii}]$\lambda$3869/[O {\sc ii}]$\lambda$3727, for
the double-peaked BLRGs, the intermediate BLRGs, and 
the single-peaked BLRGs, respectively.
The mean and median values of each flux ratio for each class of BLRG
are given in Table 2.
Figure 2 and Table 2 clearly show that the double-peaked BLRGs exhibit
larger ratios of [O {\sc i}]$\lambda$6300/[O {\sc iii}]$\lambda$5007
than the single-peaked BLRGs.
In order to examine whether or not this difference is statistically real,
we apply the Kolmogorov-Smirnov (KS) statistical test.
The KS test results in a probability that the observed frequency
distributions of [O {\sc i}]$\lambda$6300/[O {\sc iii}]$\lambda$5007
of the double-peak BLRGs and the single-peak BLRGs originate in 
the same underlying population of 1.21 $\times$ 10$^{-5}$.
Thus, we conclude that the flux ratio of 
[O {\sc i}]$\lambda$6300/[O {\sc iii}]$\lambda$5007 of the double-peaked
BLRGs is statistically larger than that of the single-peaked BLRGs.
Note that this result is consistent with the earlier remark of EH94.
Although the frequency distribution of the flux ratio of
[O {\sc i}]$\lambda$6300/[O {\sc iii}]$\lambda$5007 for the
intermediate BLRGs appears to resemble that for the single-peaked BLRGs
and to be different from that for the double-peaked BLRGs,
it is not conclusive since the number of the sample is too small.

We cannot conclude whether or not there is any statistical difference
in the flux ratios of 
[O {\sc ii}]$\lambda$3727/[O {\sc iii}]$\lambda$5007,
[S {\sc ii}]$\lambda \lambda$6717,6731/[O {\sc iii}]$\lambda$5007 and
[Ne {\sc iii}]$\lambda$3869/[O {\sc ii}]$\lambda$3727 among
the classes of BLRGs, because the number of the objects for which these 
flux ratios have been measured is too small.
We mention, however, that the flux ratio of 
[O {\sc ii}]$\lambda$3727/[O {\sc iii}]$\lambda$5007 appears to show
the same tendency as the flux ratio of 
[O {\sc i}]$\lambda$6300/[O {\sc iii}]$\lambda$5007, i.e., the 
double-peaked BLRGs exhibit larger flux ratios than the single-peaked BLRGs.
Note that we do not find significant difference in the flux ratio of
[S {\sc ii}]$\lambda \lambda$6717,6731/[O {\sc iii}]$\lambda$5007 between
the double-peaked BLRGs and the single-peaked BLRGs.
This seems rather inconsistent with the result of EH94 that 
the disk-like emitters (in the definition of EH94) present
larger equivalent width of [S {\sc ii}]$\lambda \lambda$6717,6731
than the other BLRGs though the equivalent width of
[O {\sc iii}]$\lambda$5007 is similar between the disk-like emitters and
the other BLRGs.

EH94 mentioned that the higher 
[O {\sc i}]$\lambda$6300/[O {\sc iii}]$\lambda$5007 ratio in 
double-peaked BLRGs is attributed to the relatively harder ionizing SED
arising from a nuclear ADAF, which illuminates an outer thin disk and
then causes the double-peaked profiles of Balmer lines 
(see also Halpern 1999).
However, there are other possibilities that make the flux ratio of
[O {\sc i}]$\lambda$6300/[O {\sc iii}]$\lambda$5007 high.
For instance, since the critical density of the [O {\sc i}]$\lambda$6300 
transition (1.8 $\times$ 10$^6$ cm$^{-3}$) is higher than that of the
[O {\sc iii}]$\lambda$5007 transition (7.0 $\times$ 10$^5$ cm$^{-3}$),
a high ratio of the [O {\sc i}]$\lambda$6300/[O {\sc iii}]$\lambda$5007 
is achieved when the typical gas density is high.
Thus the difference in the flux ratio of
[O {\sc i}]$\lambda$6300/[O {\sc iii}]$\lambda$5007 between
the double-peaked BLRGs and the single-peaked BLRGs may be attributed
to the difference in the typical gas density in NLRs.
The lower ionization parameter (i.e., the ratio of the ionizing photon density
to the hydrogen density) in NLRs of the double-peaked BLRGs may be also 
responsible for the observed difference in the narrow emission-line flux 
ratios.
These possibilities would not be discriminated without
photoionization model calculations taking account of the SED difference 
predicted by the illuminated-disk model, which are presented in the 
following section.

\section{PHOTOIONIZATION MODEL CALCULATIONS}

\begin{figure*}
\epsscale{1.0}
\plotone{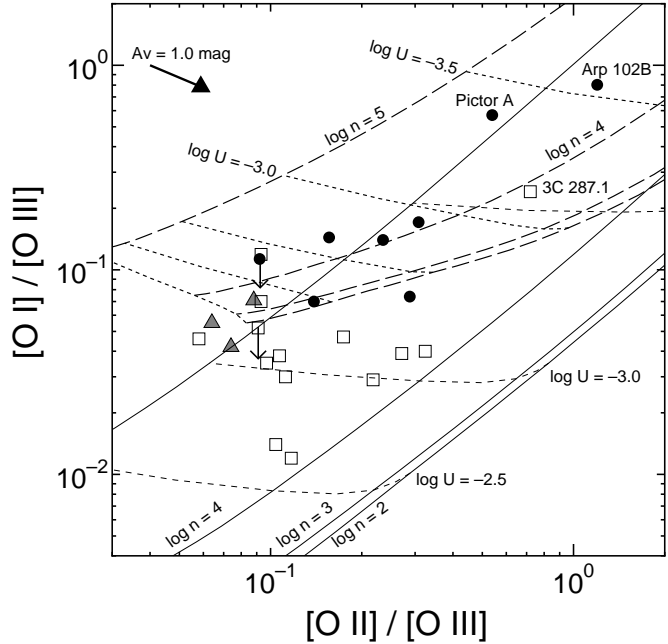}
\caption{
Diagram of [O {\sc i}]$\lambda$6300/[O {\sc iii}]$\lambda$5007 versus
[O {\sc ii}]$\lambda$3727/[O {\sc iii}]$\lambda$5007.
The observations compiled from the literature are shown by 
filled circles, gray triangles, and open squares, which denote the
double-peaked BLRGs, the intermediate BLRGs, and the single-peaked
BLRGs, respectively.
The marks with an arrow denote upper-limit data.
The results of our photoionization model calculations are shown by
solid and dashed lines, which denote the models using the 
``SED with BBB'' and the ``SED without BBB'', respectively.
The data points will move on the diagram as shown by the thick arrow 
if an extinction correction of $A_V$ = 1.0 mag is applied.
\label{fig4}}
\end{figure*}

 \subsection{Method}

As described in section 1, the illuminated-disk model hypothesizes that 
the double-peaked BLRGs harbor an ADAF in their nucleus while
the single-peaked BLRGs do not.
In order to investigate the effect of this difference on the 
gas in the NLRs and on the narrow emission-line flux ratios,
we perform photoionization model calculations using the publicly available
code $Cloudy$ version 94.00 (Ferland 1997, 2000).
Here we assume uniform density gas clouds with a plane-parallel geometry.
The parameters for the calculations are 
(I) the hydrogen density of a cloud ($n_{\rm H}$),
(II) the ionization parameter ($U$),
(III) the chemical composition of the gas, and
(IV) the shape of the SED of the input continuum radiation.
We perform several model runs covering the following ranges of parameters:
10$^{2.0}$ cm$^{-3}$ $\leq$ $n_{\rm H}$ $\leq$ 10$^{5.0}$ cm$^{-3}$ and
10$^{-4.0}$ $\leq$ $U$ $\leq$ 10$^{-1.5}$.
We set the gas-phase elemental abundances to be the solar ones.
The adopted solar abundances relative to hydrogen are taken from 
Grevesse \& Anders (1989) with extensions by Grevesse \& Noels (1993).
For simplification, dust grains in NLRs are not taken into account
in our calculations.

We prepare the two SED templates for the input continuum radiation. 
One is a typical SED for an optically-thick, geometrically-thin 
accretion disk. This is characterized by a strong BBB
in the wavelength range of UV--to--X-ray.
We adopt the empirically constructed SED (see 
Nagao, Murayama, \& Taniguchi 2001a) for this type of SED template (hereafter
``SED with BBB''). This SED is described by the following function:
\begin{equation}
f_{\nu} = \nu^{\alpha_{{\rm uv}}} \exp(-\frac{h\nu}{kT_{{\rm BB}}}) \exp
(-\frac{kT_{{\rm IR}}}{h\nu}) + a\nu^{\alpha_{{\rm x}}}
\end{equation}
(see Ferland 1997; Nagao et al. 2001a).
Here the following parameter set is adopted (Nagao et al. 2001a):
(i) the infrared cutoff of the big blue bump component, $kT_{\rm IR}$ = 
  0.01 Ryd,
(ii) the slope of the low-energy side of the big blue bump, 
  $\alpha_{\rm uv}$ = --0.5,
(iii) the UV--to--X-ray spectral index, $\alpha_{\rm ox}$ = --1.35,
(iv) the slope of the X-ray power-law continuum, $\alpha_{\rm x}$ = 
  --0.85, and
(v) the characteristic temperature of the big blue bump, $T_{\rm BB}$
  = 490,000 K\footnote{This temperature is adopted to reproduce the 
  observed soft X-ray index of nearby AGNs measured by {\it ROSAT}.
  See Nagao et al. (2001a) for more details.}.
Note that the parameter $a$ in the equation (1) is determined from
the adopted value of $\alpha_{\rm ox}$.
The last term in equation (1) is not extrapolated below 1.36 eV or
above 100 keV. Below 1.36 eV the last term is simply set to zero.
Above 100 keV the continuum is assumed to fall off as $\nu^{-3}$.
The other template is a typical SED for an optically-thin ADAF.
The SEDs produced by the optically-thin ADAF are expected to exhibit 
no strong BBB.
Recently, many efforts have been made to estimate the SED generated by ADAFs
theoretically (e.g., Narayan \& Yi 1995; Chen, Abramowicz, \& Lasota 1997;
Narayan, Kato, Honma 1997; Manmoto, Mineshige, \& Kusunose 1997; 
Kurpiewski \& Jaroszy\'{n}ski 1999; Quataert \& Narayan 1999;
Manmoto 2000; Kino, Kaburaki, \& Yamazaki 2000). 
There are, accordingly, various shapes of the calculated SEDs.
Among them, we follow the SED presented by Kurpiewski \& Jaroszy\'{n}ski (1999)
for the template (hereafter ``SED without BBB'').
This SED exhibits no BBB component and is roughly described by
single power-law continua in the range of 10$^{12}$ Hz to 10$^{20}$ Hz.
We choose the photon index of $\alpha$ = --0.89 ($f_{\nu} = \nu^{\alpha}$), 
which is predicted for the case that
a non-rotating black hole is assumed (see Kurpiewski \& Jaroszy\'{n}ski 1999).
This SED is expressed by the simple power-law spectra with the exponential
cutoffs at 10$^{-4.0}$ Ryd and 10$^{4.0}$ Ryd.
In this way, we have prepared the two SED templates (see Figure 3).
Note that these SED templates may be too simple to predict accurate
emission-line spectra radiated from NLRs in BLRGs.
However, these rough SEDs can be used to investigate the effect of the
presence or absence of the BBB component, which is the main
difference in the SED between the accretion disk with/without an ADAF.

The calculations are stopped when the temperature falls to 3000 K, below
which the gas does not contribute significantly to the observed
optical emission-line spectra.

 \subsection{Results of Model Calculations}

\begin{figure*}
\epsscale{0.9}
\plotone{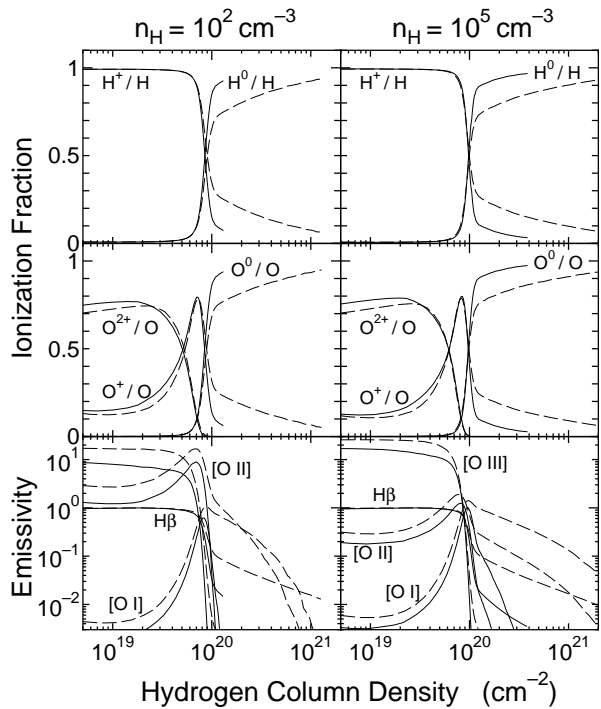}
\caption{
(Top) Ionization fractions of hydrogen as functions of depth
into the nebula for the cases of $n_{\rm H}$ = 10$^{2.0}$ cm$^{-3}$ (left)
and 10$^{5.0}$ cm$^{-3}$ (right). Here $U$ = $10^{-3.0}$ is adopted.
The solid and dashed lines denote the models adopting the
``SED with BBB'' and the ``SED without BBB'', respectively.
(Middle) Same as the top panels, but for oxygen.
(Bottom) Volume emissivities of the emission lines of
H$\beta$, [O {\sc i}]$\lambda$6300, [O {\sc ii}]$\lambda$3727, and
[O {\sc iii}]$\lambda$5007 as functions of depth into the nebula.
They are normalized by the H$\beta$ volume emissivity at the
fully-ionized region, in which the H$\beta$ volume emissivity do not
vary significantly.
\label{fig5}}
\end{figure*}

We show the results of the model calculations and compare them with the
observations in Figure 4, which is a diagram of
[O {\sc i}]$\lambda$6300/[O {\sc iii}]$\lambda$5007 versus
[O {\sc ii}]$\lambda$3727/[O {\sc iii}]$\lambda$5007.
This diagnostic diagram has been frequently used to discuss physical 
properties of gas in ionized regions (e.g., 
Baldwin, Phillips, \& Terlevich 1981).
As mentioned at section 2.2, the plotted data are
not corrected for dust extinction.
The data points will move on Figure 4 as shown by the thick arrow 
if the extinction correction of $A_V$ = 1.0 mag is applied.
Note that the direction of the extinction correction is 
perpendicular to the direction of the difference between 
the data of the double-peak and single-peak BLRGs.
This suggests that the difference in the narrow emission-line flux ratios
between the double-peaked and single-peaked BLRGs is
not caused mainly by the difference in the degree of the dust extinction.
Although the data points are clearly separated between the double-peaked
and single-peaked BLRGs in Figure 4, there is an exceptional
object, 3C 287.1; this object is classified as a single-peaked BLRG
while its locus on Figure 4 is far from the
other single-peaked BLRGs. This object, on the contrary, appears to
belong to the class of the double-peaked BLRG in terms of the
narrow emission-line flux ratios.
We speculate that this is due to the misclassification of 3C 287.1,
because the red displaced peak of H$\alpha$ is clearly seen
between [N {\sc ii}]$\lambda$6583 and [S {\sc ii}]$\lambda$6717
in the optical spectra of this object, presented by EH94.
Further observation for 3C 287.1 is necessary to examine this speculation.

As shown in Figure 4, the models adopting the SED without BBB predict a higher
[O {\sc i}]$\lambda$6300/[O {\sc iii}]$\lambda$5007 ratio by an order 
of magnitude than the models adopting the SED with BBB.
This tendency is consistent with the observations; i.e., 
the observed data of the 
single-peaked BLRGs are explained by the models adopting the
SED with BBB in the ranges of $U$ $\sim$ $10^{-3.0}$ and 
10$^{4.0}$ cm$^{-3}$ $\lesssim$ $n_{\rm H}$ $\lesssim$ 10$^{5.0}$ cm$^{-3}$
while the observed data of the double-peak BLRGs appear to be explained
introducing the models adopting the SED without BBB.

The larger flux ratios of 
[O {\sc i}]$\lambda$6300/[O {\sc iii}]$\lambda$5007 
in the models adopting the SED without BBB are attributed to the fact that 
the SED without BBB has harder spectra, which create larger partially-ionized 
regions in NLRs. To see the effect of the SED shape on the ionization 
structure of gas in NLRs, we show the ionization fractions of hydrogen 
and oxygen as functions of depth into the nebula for the cases of 
$n_{\rm H}$ = 10$^{2.0}$ cm$^{-3}$ and 10$^{5.0}$ cm$^{-3}$ in Figure 5.
Here $U$ = $10^{-3.0}$ is adopted.
As shown in the top panels of Figure 5, partially ionized regions become
large when the SED without BBB is adopted, compared to the case that
the SED with BBB is adopted.
Accordingly, the volume emissivities of low-ionization emission lines
such as [O {\sc i}]$\lambda$6300 and [O {\sc ii}]$\lambda$3727 are
higher at larger radii from the ionization source in the case that
the SED without BBB is adopted, as shown in the bottom panels of Figure 5. 
Note that the ionization structure in the nebula is almost independent
of the gas density as shown in Figure 5, though the volume emissivities of
the forbidden emission lines depend on the density.

\begin{figure*}
\epsscale{1.8}
\plotone{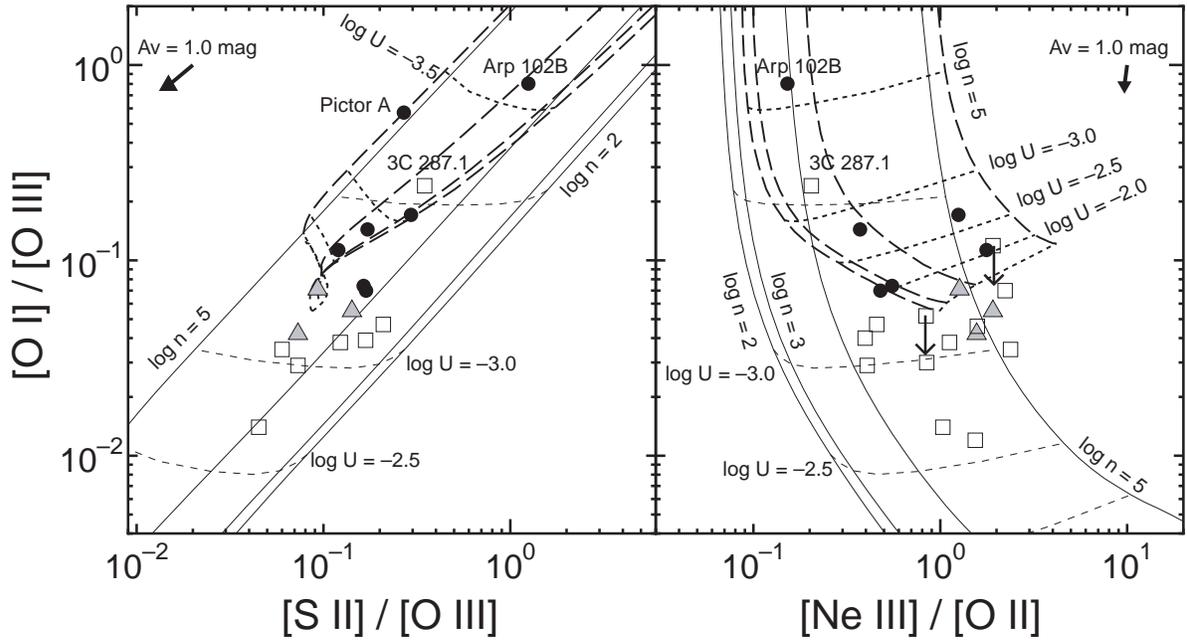}
\caption{
Diagram of [O {\sc i}]$\lambda$6300/[O {\sc iii}]$\lambda$5007 versus
[S {\sc ii}]$\lambda \lambda$6717,6731/[O {\sc iii}]$\lambda$5007 (left)
and that of [O {\sc i}]$\lambda$6300/[O {\sc iii}]$\lambda$5007 versus
[Ne {\sc iii}]$\lambda$3869/[O {\sc ii}]$\lambda$3727 (right).
The marks and the lines are the same as those in Figure 4.
\label{fig6}}
\end{figure*}

We also compare the observations with the model predictions in other 
diagnostic diagrams, which are presented in Figure 6. They are diagrams of
[O {\sc i}]$\lambda$6300/[O {\sc iii}]$\lambda$5007 versus
[S {\sc ii}]$\lambda \lambda$6717,6731/[O {\sc iii}]$\lambda$5007 and
[O {\sc i}]$\lambda$6300/[O {\sc iii}]$\lambda$5007 versus
[Ne {\sc iii}]$\lambda$3869/[O {\sc ii}]$\lambda$3727.
Similarly to the trend in Figure 4, the data of the double-peaked BLRGs and
the single-peaked BLRGs are also well separated in these diagnostic diagrams.
This difference in the location of the data on the two diagrams is well
explained by the idea that the gas in NLRs of the single-peaked BLRGs
is photoionized by the SED with BBB while that of the double-peaked BLRGs
is photoionized by the SED without BBB, which is expected 
in the framework of the illuminated-disk model.
It is noted that 3C 287.1 is again located far from the other
single-peaked BLRGs but beside the double-peaked BLRGs in the two diagrams
shown in Figure 6.

\section{Discussion}

\begin{figure*}
\epsscale{1.8}
\plotone{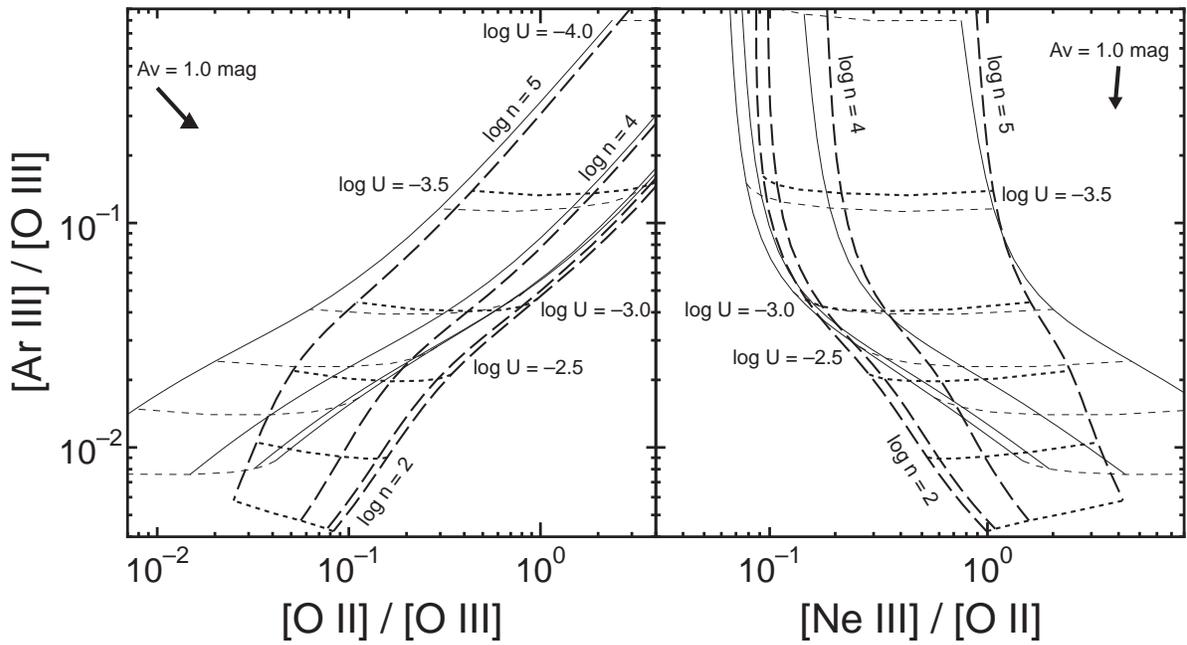}
\caption{
Diagram of [Ar {\sc iii}]$\lambda$7136/[O {\sc iii}]$\lambda$5007 versus
[O {\sc ii}]$\lambda$3727/[O {\sc iii}]$\lambda$5007 (left) and
that of [Ar {\sc iii}]$\lambda$7136/[O {\sc iii}]$\lambda$5007 versus
[Ne {\sc iii}]$\lambda$7136/[O {\sc ii}]$\lambda$3727 (right).
The lines are the same as those in Figure 4.
\label{fig7}}
\end{figure*}

As presented in section 4.2, the observed difference in the narrow 
emission-line flux ratios can be explained consistently by the 
photoionization models if we take account of the SED difference between 
the double-peaked and the single-peaked BLRGs, which is predicted by the 
illuminated-disk model.
However, the SED without BBB may not be a unique way to explain 
the difference in the observed emission-line flux ratios between 
the double-peaked and the single-peaked BLRGs. 
For instance, systematic differences in the physical properties of
gas in NLRs could be the origin of the difference in the narrow 
emission-line flux ratios.
As mentioned in section 3, the difference in the flux ratio of 
[O {\sc i}]$\lambda$6300/[O {\sc iii}]$\lambda$5007 can be also explained 
if the gas density in NLRs is systematically different between the
two populations of BLRGs.
This possibility is, however, rejected as we compare the observations
with the model predictions in the diagnostic diagrams shown in 
Figures 4 and 6.
These diagnostic diagrams suggest that there is no difference in the 
gas density between the double-peaked and single-peaked BLRGs
if we adopted only the models with the SED with BBB
(10$^{4.0}$ cm$^{-3}$ $\lesssim$ $n_{\rm H}$ $\lesssim$ 10$^{5.0}$ cm$^{-3}$).
This means that the observed difference in the flux ratio of
[O {\sc i}]$\lambda$6300/[O {\sc iii}]$\lambda$5007 is not caused only by
a difference in the gas density in NLRs.

There is another possibility which can explain the observed difference 
in the narrow emission-line flux ratios without introducing the SED 
difference. 
Since the effect of the SED difference and the varying $U$ sequence are
degenerated as displayed in Figures 4 and 6, 
a systematic difference in the ionization parameter 
of gas in NLRs between the double-peaked and the single-peaked BLRGs can be 
also responsible for the difference in the emission-line flux ratios.
Note that Pictor A and Arp 102B, which are double-peaked BLRGs, exhibit
too strong low-ionization emission lines to be explained only by the 
effect of the SED difference. 
Here it should be noted that ADAFs are expected to radiate a small 
number of ionizing photons compared to the standard $\alpha$ disks, 
which leads to a small ionization parameter of the gas in NLRs 
if the gas density and the distance from the 
nucleus of the NLRs are independent of the type of the accretion flow 
in their nuclei.
EH94 reported that the double-peaked BLRGs exhibit weaker
non-stellar continuum radiation in optical spectra than
the single-peaked BLRGs.
Therefore, the difference in the narrow emission-line flux ratios 
between the double-peaked and the single-peaked BLRGs may be attributed to
the effects of both the SED shape and the ionization parameter.
Unfortunately, the current information about the narrow emission-line 
spectra of BLRGs is insufficient to examine the contribution of
these two effects.
In order to investigate this issue, future observations of optical 
spectra with sufficient quality (i.e., measurable rather faint emission 
lines) for a large sample of BLRGs is crucially necessary.
For the sake of demonstration, we show two diagnostic diagrams
which are possibly useful to solve the degeneracy of the effects of
the SED shape and of the ionization parameter in Figure 7,
in which a diagram of
[Ar {\sc iii}]$\lambda$7136/[O {\sc iii}]$\lambda$5007 versus
[O {\sc ii}]$\lambda$3727/[O {\sc iii}]$\lambda$5007 and a diagram of 
[Ar {\sc iii}]$\lambda$7136/[O {\sc iii}]$\lambda$5007 versus
[Ne {\sc iii}]$\lambda$3869/[O {\sc ii}]$\lambda$3727 are displayed.
Since the flux ratio of 
[Ar {\sc iii}]$\lambda$7136/[O {\sc iii}]$\lambda$5007
is almost independent of the shape of the SED especially in the range of
$U \lesssim 10^{-2.5}$, these diagnostic diagrams can be used to determine 
the ionization parameter. Then we will be able to discuss the effect of 
the SED shape solely, taking account of other information of the 
emission-line spectra.

The biconical radial outflow of gas in BLRs has been often discussed
as the possible origin of the double-peaked Balmer lines as well as
the illuminated-disk model (e.g., Zheng, Binette, \& Sulentic 1990; 
Zheng et al. 1991). Here we discuss whether or not the properties of
NLR emission presented in this paper constrain this radial outflow 
model. Since this model does not require any special accretion
mechanism to generate the double-peaked Balmer lines, there is no 
reason to introduce the difference in the nuclear ionizing radiation
between the double-peaked and the single-peaked BLRGs.
However, narrow emission-line flux ratios would be different between
the two groups of the BLRG 
if the outflow in BLRs of double-peaked BLRGs affects the outer regions
and then causes a shock wave in NLRs.
Since shocks generate a large partially-ionized region in the gas,
strong low-ionization emission lines such as
[O {\sc i}]$\lambda$6300 arise from such shock-heated gas
(Mouri, Kawara, \& Taniguchi 2000 and reference therein), 
which seems consistent
with the fact that double-peaked BLRGs exhibit stronger
low-ionization emission lines than single-peaked BLRGs.
Note that Stauffer, Schild, \& Keel (1983) mentioned that
the strong low-ionization emission lines exhibited in the spectrum of
Arp 102B, a prototype of double-peaked BLRG, may be attributed
to shock excitation.
In order to examine whether or not the shock heating by the outflowing
material of the double-peaked BLRGs is responsible
for the difference in the narrow emission-line flux ratios between
the double-peaked and the single-peaked BLRGs, we plot the predictions
of the shock models presented by Dopita \& Sutherland (1995)
on the diagnostic diagrams of 
[O {\sc i}]$\lambda$6300/[O {\sc iii}]$\lambda$5007 versus
[O {\sc ii}]$\lambda$3727/[O {\sc iii}]$\lambda$5007,
[S {\sc ii}]$\lambda \lambda$6717,6731/[O {\sc iii}]$\lambda$5007, and
[Ne {\sc iii}]$\lambda$3869/[O {\sc ii}]$\lambda$3727 (Figure 8).
The shock models appear to predict too large flux ratios of
[O {\sc ii}]$\lambda$3727/[O {\sc iii}]$\lambda$5007 and too small
flux ratios of [Ne {\sc iii}]$\lambda$3869/[O {\sc ii}]$\lambda$3727
compared to the data of the double-peaked BLRGs.
These results suggest that the difference in the narrow emission-line
flux ratios is not caused by the shock-heated gas in NLRs.
Therefore, the biconical radial outflow model seems less reasonable
for the double-peaked Balmer lines.
Note, however, that this result is not conclusive because we
cannot exclude the possibility that the NLR emission of the 
double-peaked BLRGs is contributed by the shock-heated gas.
For instance, the flux ratio of 
[Ne {\sc iii}]$\lambda$3869/[O {\sc ii}]$\lambda$3727
may be possibly lower in the double-peaked BLRGs than in the 
single-peaked BLRGs (see Table 2), although it cannot be examined 
statistically due to the small number of objects.

\begin{figure*}
\epsscale{2.0}
\plotone{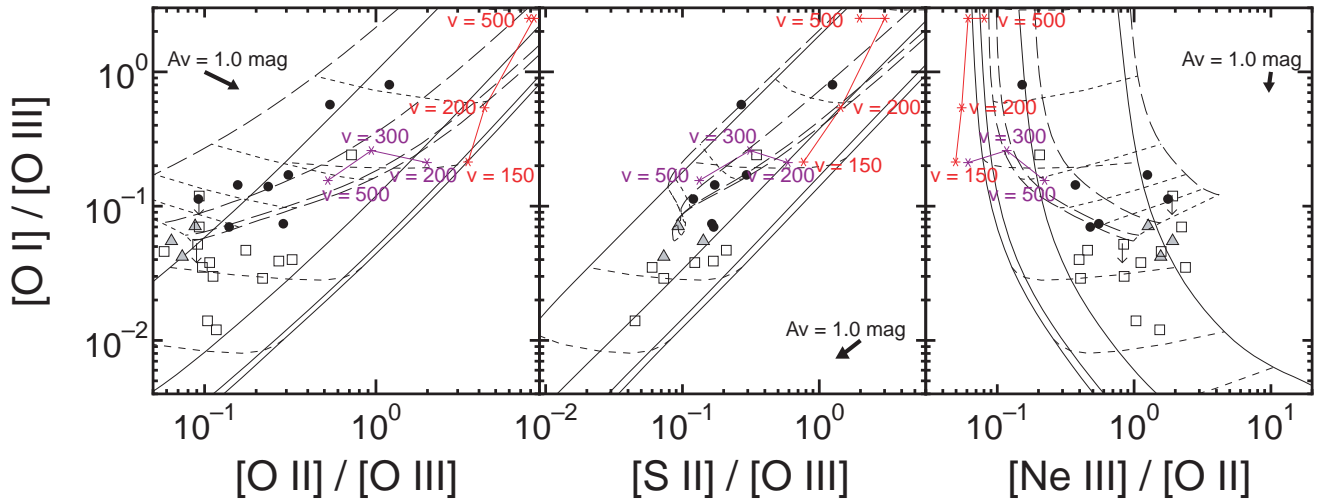}
\caption{
Diagrams of [O {\sc i}]$\lambda$6300/[O {\sc iii}]$\lambda$5007 versus
[O {\sc ii}]$\lambda$3727/[O {\sc iii}]$\lambda$5007 (left),
[S {\sc ii}]$\lambda \lambda$6717,6731/[O {\sc iii}]$\lambda$5007 (middle),
and [Ne {\sc iii}]$\lambda$3869/[O {\sc ii}]$\lambda$3727 (right).
The marks and the lines written in black are the same as in Figure 4.
The predictions of the shock model (Dopita \& Sutherland 1995) are
plotted on the diagrams by the red and purple lines.
The red line denotes the pure-shock models 
(150 km s$^{-1}$ $\leq$ $v_{\rm shock}$ $\leq$ 500 km s$^{-1}$) and
the purple line denotes the models in which the effect of precursor is
considered (200 km s$^{-1}$ $\leq$ $v_{\rm shock}$ $\leq$ 500 km s$^{-1}$).
The chemical composition of the solar abundances and the magnetic parameter
of $B/\sqrt{n}$ = 2 $\mu$G cm$^{2/3}$ are assumed in these models.
\label{fig8}}
\end{figure*}

Then, which emission-line flux ratio can be used to discriminate between
shocks and photoionization? The flux ratio of 
[O {\sc iii}]$\lambda$4363/[O {\sc iii}]$\lambda$5007, which is one of 
temperature indicators, has been often regarded as a powerful tool to 
discriminate between shocks and photoionization (e.g., 
Koski \& Osterbrock 1976; Heckman 1980; Ferland \& Netzer 1983; 
Rose \& Cecil 1983; Keel \& Miller 1983; Rose \& Tripicco 1984).
However, Nagao, Murayama, \& Taniguchi (2001b) pointed out that high ratios
of [O {\sc iii}]$\lambda$4363/[O {\sc iii}]$\lambda$5007 do not always mean 
the shock excitation. They showed that the photoionization models in which
high-density gas clouds are taken into account can explain the observed high 
ratios of [O {\sc iii}]$\lambda$4363/[O {\sc iii}]$\lambda$5007 (see also 
Nagao et al. 2001c). Therefore, we focus on the emission-line flux ratio of 
[S {\sc iii}]$\lambda$9532/[S {\sc ii}]$\lambda \lambda$6717,6731. This flux 
ratio is expected to be low when the gas is excited by shocks, because
S$^{2+}$ ions cool predominantly by emission of UV lines due to the higher 
temperature and thus the [S {\sc iii}]$\lambda$9532 line becomes weak when 
the shock contributes to the excitation significantly (e.g., Dopita 1977; 
D\'{\i}az, Pagel, \& Wilson 1985; D\'{\i}az, Terlevich, \& Pagel 1985;
Bonatto, Bica, \& Alloin 1989;
Kirhakos \& Phillips 1989; Simpson et al. 1996). 
Note that this tool can work even in the case that the 
gas density is rather high, being different from the flux ratio of 
[O {\sc iii}]$\lambda$4363/[O {\sc iii}]$\lambda$5007.
We compare the photoionization models adopting the SED with/without BBB
with the shock models on the diagrams of 
[S {\sc iii}]$\lambda$9532/[S {\sc ii}]$\lambda \lambda$6717,6731 versus
[O {\sc i}]$\lambda$6300/[O {\sc iii}]$\lambda$5007, 
[O {\sc ii}]$\lambda$3727/[O {\sc iii}]$\lambda$5007, and 
[S {\sc ii}]$\lambda \lambda$6717,6731/[O {\sc iii}]$\lambda$5007 (Figure 9).
It is clearly shown that these diagnostic diagrams can be used to
discriminate between shock and photoionization. Further observations of
the [S {\sc iii}]$\lambda$9532 emission will make it clear whether or not 
the observed differences in the narrow emission-line flux ratios between the 
double-peaked and single-peaked BLRGs are attributed to the shock excitation.

Now we mention the intermediate BLRGs. On the diagnostic diagrams 
presented in Figures 4 and 6, the data of the intermediate BLRGs appear 
to concentrate near the data of the single-peaked BLRGs and to be 
separated from the data of the double-peaked BLRGs. Moreover, it seems hard
to explain the observed emission-line flux ratios of the intermediate 
BLRGs by the photoionization models adopting the SED without BBB, since 
unreasonably large ionization parameters ($U > 10^{-1.5}$) are required (see 
Figures 4 and 6). This suggests that the gas clouds in NLRs of the intermediate
BLRGs are photoionized by the ionizing continuum with BBB, which is expected 
in the case that the accretion disk in their nuclei is characterized by a
standard optically-thick (and geometrically-thin) disk. We thus speculate
that an ADAF occupies only a very small part of the inner accretion disk
in the intermediate BLRGs. If this is the case, the thin disk can emit
a substantial BBB emission (see Gammie, Blandford, \& Narayan 1999). 
However, since the number of the intermediate BLRGs whose NLR emission 
is investigated is too small, further observations are necessary for the 
discussion of this issue in detail.

Recently, it has been reported that some of LINERs also show the 
double-peaked Balmer lines (e.g., Storchi-Bergmann et al. 1993;
Bower et al. 1996; Ho et al. 2000; Shields et al. 2001; 
see also Barth et al. 2001; Eracleous \& Halpern 2001).
Since some lines of evidence that the LINERs harbor an ADAF in their nucleus,
the reason why the LINERs show lower ionization parameters than 
other AGNs such as Seyfert nuclei may be the harder SED of the ionizing
continuum radiation which is caused by the ADAFs.
To investigate this issue, detailed photoionization model calculations
for the NLR emission of LINERs will be necessary.
We will investigate this issue in a subsequent paper.

\begin{figure*}
\epsscale{2.0}
\plotone{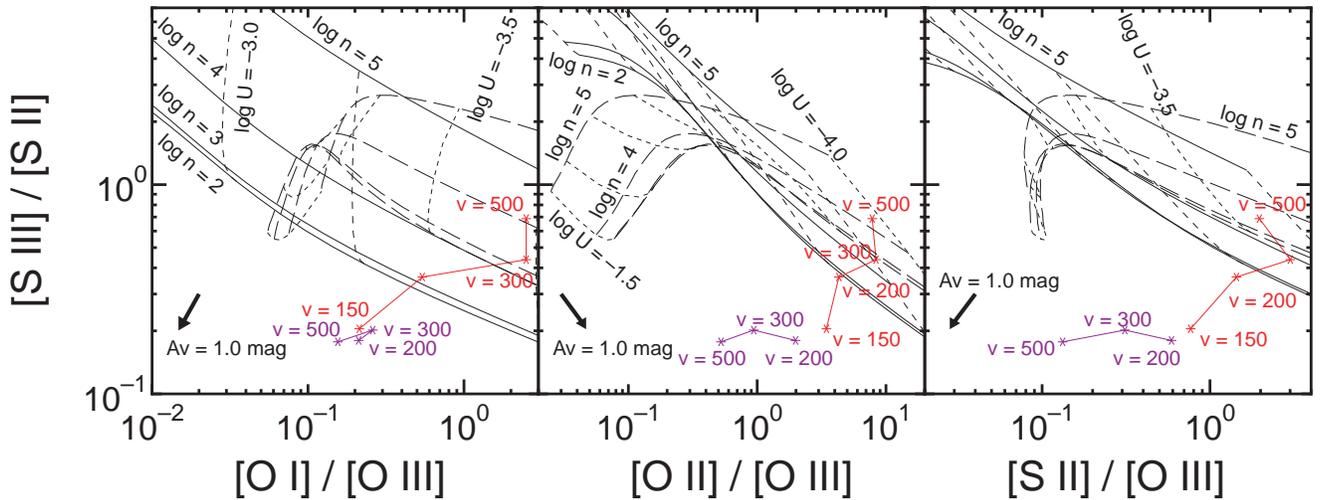}
\caption{
Diagrams of [S {\sc iii}]$\lambda$9532/[S {\sc ii}]$\lambda \lambda$6717,6731
versus [O {\sc i}]$\lambda$6300/[O {\sc iii}]$\lambda$5007 (left),
[O {\sc ii}]$\lambda$3727/[O {\sc iii}]$\lambda$5007 (middle), and
[S {\sc ii}]$\lambda \lambda$6717,6731/[O {\sc iii}]$\lambda$5007 (right).
The lines are the same as in Figure 8.
The plotted shock models are the same as presented in Figure 8.
\label{fig9}}
\end{figure*}

\section{CONCLUSION}

Based on the data compilation of narrow emission-line flux ratios
from the literature, we find that the data of the double-peaked BLRGs and
those of the single-peaked BLRGs are clearly separated on some
diagnostic diagrams which consist of the narrow emission-line flux ratios of
[O {\sc i}]$\lambda$6300/[O {\sc iii}]$\lambda$5007,
[O {\sc ii}]$\lambda$3727/[O {\sc iii}]$\lambda$5007,
[S {\sc i}]$\lambda \lambda$6717,6731/[O {\sc iii}]$\lambda$5007, and
[Ne {\sc iii}]$\lambda$3869/[O {\sc ii}]$\lambda$3727.
Our photoionization model calculations show that the data of the 
single-peaked BLRGs are well described by the models adopting the 
SED with BBB in the ranges of the parameters of $U$ $\sim$ $10^{-3.0}$ and 
10$^{4.0}$ cm$^{-3}$ $\lesssim$ $n_{\rm H}$ $\lesssim$ 10$^{5.0}$ cm$^{-3}$
while the data of the double-peaked BLRGs are consistent with the models 
adopting the SED without BBB.
This is consistent with the illuminated-disk model that predicts 
the existence of an ADAF, which illuminates the outer thin disk
and causes the double-peaked emission lines, in a nucleus of a 
BLRG with the double-peaked Balmer lines.

\acknowledgments

We would like to thank Gary Ferland for providing his code $Cloudy$ 
to the public. Charles R. Lawrence kindly provided us some useful 
information about spectroscopic properties of 3C 390.3.
We also thank Naohisa Anabuki for fruitful discussion.
The anonymous referee and Motoki Kino gave us useful comments.
We also acknowledge Yuko Kakazu for her assistance.
YS is supported by a Research Fellowship from the Japan Society for the 
Promotion of Science for Young Scientists.
This work was financially supported in part by Grant-in-Aids for the 
Scientific Research (Nos. 10044052, 10304013, and 13740122) of the 
Japanese Ministry of Education, Culture, Sports, Science, and Technology.



\begin{deluxetable}{lcccccc}
\tablenum{2}
\tablecaption{Mean and Median Emission-Line Flux Ratios \label{tbl-2}}
\tablehead{
\colhead{} &
\multicolumn{2}{l}{Double-Peaked BLRG} &
\multicolumn{2}{l}{Intermediate BLRG} &
\multicolumn{2}{l}{Single-Peaked BLRG\tablenotemark{a}} \\
\colhead{} &
\colhead{Mean} &
\colhead{Median} & 
\colhead{Mean} & 
\colhead{Median} & 
\colhead{Mean} & 
\colhead{Median} 
}
\startdata
[O {\sc i}]/[O {\sc iii}]   & 
   0.216 & 0.132 & 0.048 & 0.052 & 0.053 & 0.044 \nl
[O {\sc ii}]/[O {\sc iii}]  & 
   0.395 & 0.288 & 0.075 & 0.074 & 0.218 & 0.114 \nl
[S {\sc ii}]/[O {\sc iii}]  & 
   0.348 & 0.172 & 0.103 & 0.093 & 0.505 & 0.143 \nl
[Ne {\sc iii}]/[O {\sc ii}] &  
   0.762 & 0.515 & 1.581 & 1.565 & 1.095 & 0.846 
\enddata
\tablenotetext{a}{The upper-limit data are not included.}
\end{deluxetable}

\begin{deluxetable}{lccccl}
\scriptsize
\tablenum{1}
\tablecaption{Compiled Data \label{tbl-1}}
\tablehead{
\colhead{Name} &
\colhead{[O {\sc i}]/[O {\sc iii}]} &
\colhead{[O {\sc ii}]/[O {\sc iii}]} & 
\colhead{[S {\sc ii}]/[O {\sc iii}]} & 
\colhead{[Ne {\sc iii}]/[O {\sc ii}]} &
\colhead{References\tablenotemark{a}} 
}
\startdata
\cutinhead{Double-Peaked BLRG}
3C 17           &\nodata & 0.603  &\nodata &\nodata & 1, 2      \nl
3C 59           & 0.070  & 0.139  & 0.169  & 0.478  & 3         \nl
IRAS 0236.6-3101& 0.108  &\nodata &\nodata &\nodata & 4         \nl
PKS 0340-37     & 0.110  &\nodata &\nodata &\nodata & 4         \nl
3C 93           & 0.077  &\nodata &\nodata &\nodata & 4         \nl
1E 0450.3-1817  & 0.731  &\nodata &\nodata &\nodata & 5         \nl
Pictor A        & 0.571  & 0.539  & 0.269  &\nodata & 1, 6, 7, 8\nl
B2 0742+31      & 0.030  &\nodata &\nodata &\nodata & 4         \nl
PKS 0857-19     & 0.186  &\nodata &\nodata &\nodata & 4         \nl
PKS 1151-34     & 0.140  & 0.235  &\nodata &\nodata & 1, 4, 9   \nl
Mrk 668         & 0.037  &\nodata &\nodata &\nodata & 4         \nl
3C 303          & 0.251  &\nodata &\nodata &\nodata & 4         \nl
3C 332          & 0.144  & 0.156  & 0.172  & 0.372  & 3         \nl
3C 382          & 0.074  & 0.288  & 0.164  & 0.552  & 2, 10     \nl
3C 390.3        & 0.113  & 0.092  & 0.120  & 1.767  & 2, 10, 11, 12\nl
Arp 102B        & 0.803  & 1.197  & 1.248  & 0.152  & 13        \nl
PKS 1739+18     & 0.154  &\nodata &\nodata &\nodata & 4         \nl
PKS 1914-45     & 0.123  &\nodata &\nodata &\nodata & 4         \nl
PKS 2300-18     & 0.171  & 0.308  & 0.294  & 1.250  & 8, 14     \nl
\cutinhead{Intermediate BLRG}
PKS 0202-76     & 0.055  &\nodata &\nodata &\nodata & 4         \nl
4C 5.38         & 0.049  &\nodata &\nodata &\nodata & 4         \nl
PKS 1232-24     & 0.017  &\nodata &\nodata &\nodata & 4         \nl
3C 227          & 0.071  & 0.088  & 0.093  & 1.268  & 1, 10, 15 \nl
4C 73.18        & 0.055  & 0.064  & 0.142  & 1.911  & 16        \nl
3C 445          & 0.042  & 0.074  & 0.073  & 1.565  & 1, 8, 10, 17\nl
\cutinhead{Single-Peaked BLRG}
B2 0110+29      & 0.019  &\nodata &\nodata &\nodata & 4         \nl
3C 48           &\nodata & 0.895  & 3.739  & 0.714  & 18, 19    \nl
PHL 1093        &$<$0.306&\nodata &\nodata &\nodata & 20        \nl
3C 61.1         & 0.047  & 0.174  & 0.209  & 0.457  & 16        \nl
PKS 0214+10     & 0.042  &\nodata &\nodata &\nodata & 4         \nl
3C 109          &\nodata & 0.093  &\nodata &\nodata & 2         \nl
3C 120          & 0.035  & 0.097  & 0.060  & 2.376  & 1, 21, 22 \nl
PKS 0736+01     &\nodata &$<$0.080&\nodata &\nodata & 1         \nl
3C 206          & 0.070  & 0.093  &\nodata & 2.222  & 4, 20     \nl
4C 9.35         & 0.036  &\nodata &\nodata &\nodata & 4         \nl
3C 234          & 0.014  & 0.104  & 0.045  & 1.032  & 2. 3      \nl
PKS 1011-282    &\nodata & 0.093  &\nodata &\nodata & 23        \nl
3C 249.1        &\nodata & 0.222  & 0.138  & 0.787  & 18, 24, 25\nl
PKS 1101-32     & 0.046  & 0.058  &\nodata & 1.578  & 4, 26     \nl
PKS 1217+02     &$<$0.119& 0.093  &\nodata & 1.915  & 20        \nl
B2 1223+25      & 0.012  & 0.117  &\nodata & 1.539  & 4, 20, 23 \nl
3C 273          &\nodata &$<$0.161& 0.148  &\nodata & 1, 18     \nl
3C 277.1        & 0.040  & 0.324  &\nodata & 0.396  & 4, 27     \nl
3C 287.1        & 0.241  & 0.720  & 0.348  & 0.204  & 3         \nl
PKS 1355-41     &\nodata & 0.116  &\nodata &\nodata & 1         \nl
PKS 1417-19     & 0.038  & 0.107  & 0.123  & 1.120  & 3, 14, 28 \nl
PKS 1421-38     & 0.078  &\nodata &\nodata &\nodata & 4         \nl
4C 37.43        & 0.029  & 0.218  & 0.073  & 0.406  & 4, 18     \nl
4C 35.37        &\nodata & 0.356  &\nodata &\nodata & 3         \nl
3C 323.1        & 0.030  & 0.112  &\nodata & 0.846  & 20, 27    \nl
B2 1719+35      & 0.060  &\nodata &\nodata &\nodata & 4         \nl
MC 1745+16      & 0.046  &\nodata &\nodata &\nodata & 4         \nl
PKS 2135-14     &$<$0.052& 0.091  &\nodata & 0.839  & 1, 20     \nl
PKS 2139-04     & 0.055  &\nodata &\nodata &\nodata & 4         \nl
PKS 2208-13     & 0.073  &\nodata &\nodata &\nodata & 4         \nl
PKS 2227-399    & 0.058  &\nodata &\nodata &\nodata & 4         \nl
PKS 2302-71     & 0.047  &\nodata &\nodata &\nodata & 4         \nl
PKS 2349-01     & 0.039  & 0.271  & 0.168  &\nodata & 3
\enddata
\tablenotetext{a}{References. --- 
                  (1)~Tadhunter et al. 1993; 
                  (2)~Yee \& Oke 1978;
                  (3)~Grandi \& Osterbrock 1978;
                  (4)~Eracleous \& Halpern 1994;
                  (5)~Stephens 1989;
                  (6)~Carswell et al. 1984;
                  (7)~Filippenko 1985;
                  (8)~Robinson et al. 1987;
                  (9)~Morganti et al. 1997;
                  (10)~Osterbrock, Koski, \& Phillips 1976;
                  (11)~Lawrence 2001 (private communication);
                  (12)~Zheng et al. 1995;
                  (13)~Stauffer, Schild, \& Keel 1983;
                  (14)~Hunstead, Murdoch, \& Shobbrook 1978;
                  (15)~Simpson et al. 1996;
                  (16)~Lawrence et al. 1996;
                  (17)~Morris \& Ward 1988;
                  (18)~Boroson \& Oke 1984;
                  (19)~Phillips 1978;
                  (20)~Baldwin 1975;
                  (21)~Baldwin et al. 1980;
                  (22)~Durret \& Bergeron 1988;
                  (23)~Boisson et al. 1994;
                  (24)~Dultzin-Hacyan 1985;
                  (25)~Richstone \& Oke 1977;
                  (26)~Maza \& Ruiz 1989;
                  (27)~Wills et al. 1993
                  (28)~Rodr\'{\i}guez-Ardila, Pastoriza, \& Donzelli 2000.
}
\end{deluxetable}

\end{document}